\documentclass[useAMS,usenatbib]{mn2e}
\usepackage{graphicx}
\usepackage{subfigure}

\newcommand{\Msun}{\mbox{$M_{\sun}$}}
\newcommand{\Mwd}{\mbox{$M_{\rm wd}$}}
\newcommand{\Rwd}{\mbox{$R_{\rm wd}$}}
\newcommand{\vff}{\mbox{$v_{\rm ff}$}}
\newcommand{\degree}{^{\scriptscriptstyle{\circ}}}
\newcommand{\gcms}{\mbox{$\rm g \, cm^{-2} \, s^{-1}$}}

\title[Compton Scattering of Fe K$\alpha$ Lines in mCVs]{Compton Scattering of 
Fe K$\alpha$ Lines in Magnetic Cataclysmic Variables}
\author[McNamara et al.]
{A. L. McNamara$^{1}$\thanks{E-mail:aimee, kuncic@physics.usyd.edu.au}, 
Z. Kuncic$^{1}$\footnotemark[1],  
K. Wu$^{1,2}$, D. K. Galloway$^{3}$\thanks{Centenary Fellow} and J. G. Cullen$^{4}$\\
$^{1}$School of Physics, University of Sydney, NSW 2006, Australia\\
$^{2}$Mullard Space Science Laboratory, University College London, Holmbury St
Mary, Surrey, RH5 6NT, UK\\
$^{3}$School of Physics, University of Melbourne, Victoria 3010, Australia\\
$^{4}$Thales, Garden Island, Cowper Wharf Rd, Potts Point, NSW 2011, Australia}

\begin{document}

\date{Accepted ... . Received ...; in original form ... }

\pagerange{\pageref{firstpage}--\pageref{lastpage}} \pubyear{2006}

\maketitle

\label{firstpage}

\begin{abstract}
Compton scattering of X-rays in the bulk flow of the accretion column in 
magnetic cataclysmic variables (mCVs) can significantly shift photon energies. 
We present Monte Carlo simulations based on a nonlinear algorithm 
demonstrating the effects of Compton scattering on the H-like, He-like and 
neutral Fe K$\alpha$ lines produced in the post-shock region of the accretion 
column. The peak line emissivities of the photons in the post-shock flow are 
taken into consideration and frequency shifts due to Doppler effects are also 
included. We find that line profiles are most distorted by Compton scattering 
effects in strongly magnetized mCVs with a low white dwarf mass and high mass 
accretion rate and which are viewed at an oblique angle with respect to the
accretion column. The resulting line profiles are most sensitive to the 
inclination angle. We have also explored the effects of modifying the 
accretion column width and using a realistic emissivity profile. We find that 
these do not have a significant overall effect on the resulting line profiles.
A comparison of our simulated line spectra with high resolution 
\textit{Chandra}/HETGS observations of the mCV GK Per indicates that a wing 
feature redward of the 6.4\,keV line may result from Compton recoil near the 
base of the accretion column.

\end{abstract}


\section{Introduction}
Magnetic cataclysmic variables (mCVs) are close interacting binaries consisting
of a magnetic white dwarf (WD) and a low mass red dwarf \citep*{Warner}. Near 
the white dwarf surface the accretion flow in mCVs is confined by the magnetic 
field of the WD and is channelled to the magnetic pole region(s) of the WD, 
forming an accretion column \citep[see][for reviews]{Warner,Cropper1990,
Wu2000, Wu03}.

Near the base of the accretion column, material in supersonic free-fall is 
brought to rest on the WD surface, forming a standing shock which heats and 
ionizes the accreting plasma. The shock temperature $T_{\rm s}$ depends mainly 
on the mass $\Mwd$ and radius $\Rwd$ of the WD and is given by 
\citep[e.g.][]{Wu2000}
\begin{equation}
  kT_{\rm s}=\frac{3}{8}\frac{GM_{\rm wd}{\mu}m_{\rm H}}{R_{\rm wd} + x_{\rm s}} \qquad ,
\label{shocktemp}
\end{equation} 
where $\mu$ is the mean molecular weight and $x_{\rm s}$ is the shock height. 
For WD masses $\Mwd \approx (0.5 - 1.0) \, \Msun$ and typical mCV parameters, 
the shock temperature is $kT_{\rm s} \approx (10-40)\, \rm keV \approx(1-4)
\times10^8$ K. The plasma in the post-shock region of the column cools by 
emitting bremsstrahlung X-rays and optical/IR cyclotron radiation 
\citep*{Lamb, King}. Since cooling occurs along the flow, the post-shock region
is stratified in density and temperature. The height of the post-shock region 
is determined by the cooling length. For a flow with only bremsstrahlung 
cooling, the shock height is given by \citep*{Wu94}
\begin{eqnarray}
 x_{\rm s} \approx 3\times 10^7 \left(\frac{\dot m}{1 \,\rm \gcms}\right)
\left(\frac{\Mwd}{0.5\, \Msun}\right)^{3/2} \nonumber \\
   \times \left(\frac{\Rwd}{10^9 \rm cm}\right)^{-3/2} {\rm cm}  
\label{shockheight}
\end{eqnarray}
where $\dot m$ is the specific mass accretion rate.
 
A plasma temperature of $kT \approx 10$ keV is sufficient to fully ionize 
elements such as argon, silicon, sulphur, aluminium or calcium. Heavier 
elements such as iron can be highly ionized, resulting in H-like Fe 
{\scriptsize{XXVI}} and He-like Fe {\scriptsize{XXV}} ions. K-shell transitions
in Fe {\scriptsize{XXVI}} and Fe {\scriptsize{XXV}} ions give rise to 
K$\alpha$ lines at 6.97\,keV and 6.675\,keV, respectively \citep*[see e.g.][]
{Wu01}. The irradiation of low ionized and neutral iron by X-rays above the Fe 
K edge produces fluorescent K$\alpha$ emission at approximately 6.4\,keV inside
the WD atmosphere, beneath the accretion column and in surrounding areas. The 
natural widths of the Fe K$\alpha$ lines are small, but the lines can be 
Doppler broadened by the bulk and thermal motions of the emitters in the 
post-shock flow. The bulk velocity immediately downstream of the shock is 
$\approx 0.25\left(G\Mwd/\Rwd \right)^{1/2}\sim 1000 \, \rm km \, s^{-1}$ for 
typical mCV parameters. Lines can also be broadened by scattering processes. 
Compton (electron) scattering is expected to be more important than resonance 
(ion) scattering for the K$\alpha$ transitions \citep*{Pozd83}. For an mCV with
specific accretion rate $\dot m \sim 10~\gcms$ the electron number density is 
$n_{\rm e} \sim 10^{16} \rm \, cm^{-3}$ for a shock heated region of thickness 
$x_{\rm s} \sim 10^{7} \rm \, cm$, giving a Thompson optical depth of $\tau 
\sim 0.1$. Thus, one in every ten photons would encounter an electron before 
leaving the post-shock region. The relative importance of Doppler shifts, 
thermal Doppler broadening and Compton scattering depends on the ionization 
structure in the post-shock flow.  

X-ray observations by \textit{Chandra}/HETGS and \textit{ASCA}/SIS have
revealed significant broadening of some Fe K$\alpha$ lines in mCVs 
\citep*{Hellier98,Hellier04}. It was suggested that Compton scattering in the 
accretion column is largely responsible for the broadening in the observed 
lines. Doppler broadening should only be significant in lines emitted close to 
the shock. The absence of Doppler shifts in the observed H-like and He-like 
lines suggests that these photons may be emitted predominantly from regions of 
lower velocity near the base of the accretion column \citep{Hellier04}. The 
observed line profiles have yet to be fully interpreted with a quantitative 
model that takes into account Compton scattering effects in a complex 
ionization structure.    

In this paper, we study the effects of Compton scattering in the accretion 
column of mCVs using a nonlinear Monte Carlo algorithm 
\citep*{Cullena, Cullenb} that self-consistently takes into account the 
density, velocity and temperature structure in the column \citep{Wu01}. The 
effects of dynamical Compton scattering are also included. In a preliminary 
investigation \citep*{Kuncic}, it was found that Fe K$\alpha$ line photons 
emitted from the dense base of the accretion column undergo multiple Compton 
scatterings and as a result, the base of the line profile is substantially 
broadened. Photons emitted near the shock can also undergo scatterings with 
hot electrons immediately downstream of the shock, as well as cold electrons in
the pre-shock flow before escaping the column. The resulting line profiles 
display a shoulder-like feature redward of the line centre. More significant 
broadening is observed when cyclotron cooling is sufficiently strong to produce
a dense, compact post-shock region \citep*[see][for example]{Wu2000}. Here, we 
make three substantial improvements to the previous study: (i) we include 
Doppler effects; (ii) the photon source regions in the post-shock column are 
determined from the ionization structure, rather than specified arbitrarily; 
and (iii) the effects of different viewing angles are fully explored. The paper
is organized as follows: the theoretical outline and geometry of the model are 
described in Section 2. Numerical results for cases where cyclotron cooling is 
negligible and when it dominates are presented and discussed in Section 3. A 
summary and conclusions are presented in Section 4.

\section{Theoretical Model}
\subsection{Physical Processes}
Line photons can undergo energy changes when scattering with electrons, 
resulting in distortions in the line profile. The energy change of a photon 
per scattering is given by \citep[e.g. see][]{Pozd83},
\begin{equation}
\frac{E'}{E} = \frac{1 - {\mu}{\beta}} { 1 - {\mu}'{\beta} + \frac{E}{ {\gamma}m_{\rm e}c^2 } (1 - {\cos}{\alpha})}
\label{Compton}
\end{equation}
where E is the initial photon energy, $\gamma m_{\rm e} c^2$ is the electron 
energy, with $\gamma = (1 - \beta^2)^{-1/2}$ and where $\beta = v/c$ includes 
both thermal and bulk motion, $\theta = \cos^{-1}{\mu}$ is the incident photon 
propagation angle measured relative to the electron's direction of motion, and 
the scattering angle is $\alpha$. The prime superscript denotes quantities 
after a scattering event. Although the energy change per scattering is 
typically small, the line profile can be broadened considerably as a result of 
multiple scatterings if the optical depth is large. 

Photons scattering with hot electrons (i.e. $kT_{\rm e} > E_{\rm c}$ where 
$T_{\rm e}$ is the electron temperature and $E_{\rm c}$ is the line centre 
energy) will gain energy, while photons scattering with cold electrons 
($kT_{\rm e} \ll E_{\rm c}$) will lose energy due to recoils. In the mCV 
context, Compton recoil can be important for photons scattering with electrons 
in the cold pre-shock flow and also near the base of the column, where
$\beta$ rapidly decreases and where the optical depth is high. From 
equation~(\ref{Compton}), the fractional energy change due to Compton recoil 
$\left(E \gg \gamma m_{\rm e}c^2 \right)$ is 
\begin{equation}
  \frac{{\Delta}E}{E} \simeq - \frac{E}{m_{\rm e}c^2}(1 - \cos\alpha) \qquad .
\end{equation}

In the post-shock accretion column in mCVs, line photons undergo thermal 
Doppler broadening as well as Doppler shifts. The bulk velocity of the 
accreting material immediately downstream of the shock is $\sim 1000 \, 
\rm km \, s^{-1}$. For all inclination angles $i$ (see Figure~\ref{geometry}), 
the bulk flow is moving away from our line of sight, so the line centre energy 
is redshifted by an amount $\Delta E/E \sim \beta \cos i$. However, since the 
bulk velocity in the post-shock flow in mCVs is always less than a few 
$\times \, 1000 \, \rm km \, s^{-1}$, giving $\beta < 10^{-2}$, Doppler shifts 
are expected to be negligible.  Thermal Doppler broadening, on the other hand, 
is expected to be of order ${\Delta}E/E \sim \sqrt{(2kT_{\rm s}/m_{\rm i}c^2)} 
\approx 0.002$ for lines emitted in the hottest regions of the post-shock flow 
(i.e. immediately downstream of the shock), where $m_{\rm i}$ is the mass of 
the ion. 

In strongly magnetized mCVs, cyclotron emission is the dominant cooling 
process. The effect of this additional electron cooling process in the 
post-shock region is to reduce the shock height and modify the density and 
temperature structure of the region \citep{Wu94}. This can enhance Compton 
scattering features in line profiles \citep{Kuncic}.

\subsection[]{Geometry of the Accretion Column}

\begin{figure}
\includegraphics[width=9.0truecm]{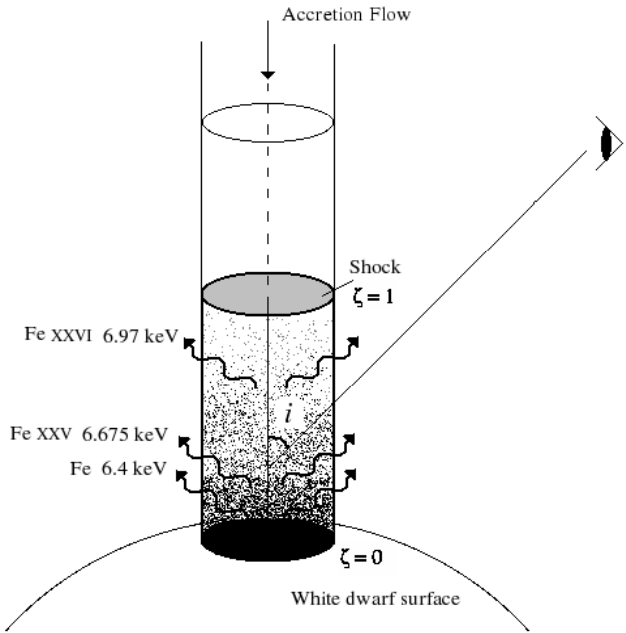}
\caption{Schematic illustration of the geometry of the magnetized white dwarf 
         accretion column showing approximate locations of Fe K$\alpha$ source 
	 regions.}
\label{geometry}
\end{figure}

The geometry of an mCV accretion column is shown in figure~\ref{geometry}.
The column is modelled as a cylinder and is divided into a shock heated
region and a cool pre-shock region. In the pre-shock region, the density and 
velocity of the accreting material are constant. The velocity of the pre-shock 
flow is approximated by the free-fall velocity at the shock height, 
$\vff(x_{\rm s}) = [G\Mwd / (\Rwd + x_{\rm s})]^{1/2}$, and the plasma is 
assumed to be cold ($kT_{\rm e} \ll 1 \rm ~keV$). The electron number density 
in the pre-shock flow is  $n_{\rm e} = \dot m/({\mu} m_{\rm H} \vff)$, where 
$\dot m$ is the specific mass accretion rate. In the post-shock region, the 
density, velocity and temperature profiles are calculated using the 
hydrodynamic solution described in \citet{Wu94}, for bremsstrahlung and 
cyclotron cooling. The parameters that determine the structure of the 
post-shock region are the WD mass $\Mwd$, the WD radius $\Rwd$, the specific 
mass accretion rate $\dot m$ and the ratio of the efficiencies of cyclotron 
cooling to bremsstrahlung cooling $\epsilon_{\rm s}$.

The accretion column is viewed from an inclination angle $i$, measured relative
to the column axis (see Fig.~\ref{geometry}). An inclination angle $i=0\degree$
corresponds to viewing the column along its axis, towards the WD, while viewing
the column from an inclination angle $i = 90\degree$ is equivalent to viewing 
the column from the side.
    
The line photons are injected into the post-shock region at a specific 
dimensionless height $\zeta \equiv {x}/{x_{\rm s}}$ at or above the WD surface,
where ${\zeta} = 0$ and ${\zeta} = 1$ correspond to the WD surface and shock 
surface, respectively (see Fig.~\ref{geometry}). For each set of mCV 
parameters, the injection point of the line corresponds to the location in the 
post-shock region where the emissivity of the specific line peaks, according to
the ionization structure determined by \citet{Wu01}. The bulk and thermal 
velocities at the line injection height $\zeta$ are used to calculate the 
Doppler shift and broadening of the lines. 

The Monte Carlo technique is used to model Compton scattering effects on 
photon propagation in the column. The distance to a tentative scattering point 
is determined using an algorithm based on a nonlinear transport 
technique \citep{Stern} which integrates the mean free path over the spatially 
varying electron density \citep{Cullena, Cullenb}. The scattering cross-section
is determined from the Klein-Nishina formula and the momentum vector at the 
scattering point is drawn from an isotropic Maxwellian distribution at the 
local temperature. A rejection algorithm is used to decide whether the 
scattering is accepted \citep{Cullenb}. For an accepted event the energy and 
momentum changes of the photon are calculated as described in \citet{Pozd83}. 
In each simulation photons are followed until they leave the column and binned 
to form a spectrum. A full description of the numerical algorithm can be found 
in \citet{Cullena}.
  
In the simulations described below, we make the following simplifying 
assumptions: a fixed number of $10^8$ photons are used to simulate each line; 
a single energy is used for each line (whereas in reality, the neutral and 
Lymann $\alpha$ transitions are doublets, with energies 6.391/6.404~keV and 
6.952/6.973~keV, and the He-like transition has both resonant, 
inter-combination and forbidden components); a fixed accretion column width 
is used and a single line injection site is used.

\section{Results and Discussion}
\begin{figure*}
  \includegraphics[width=5.5truecm]{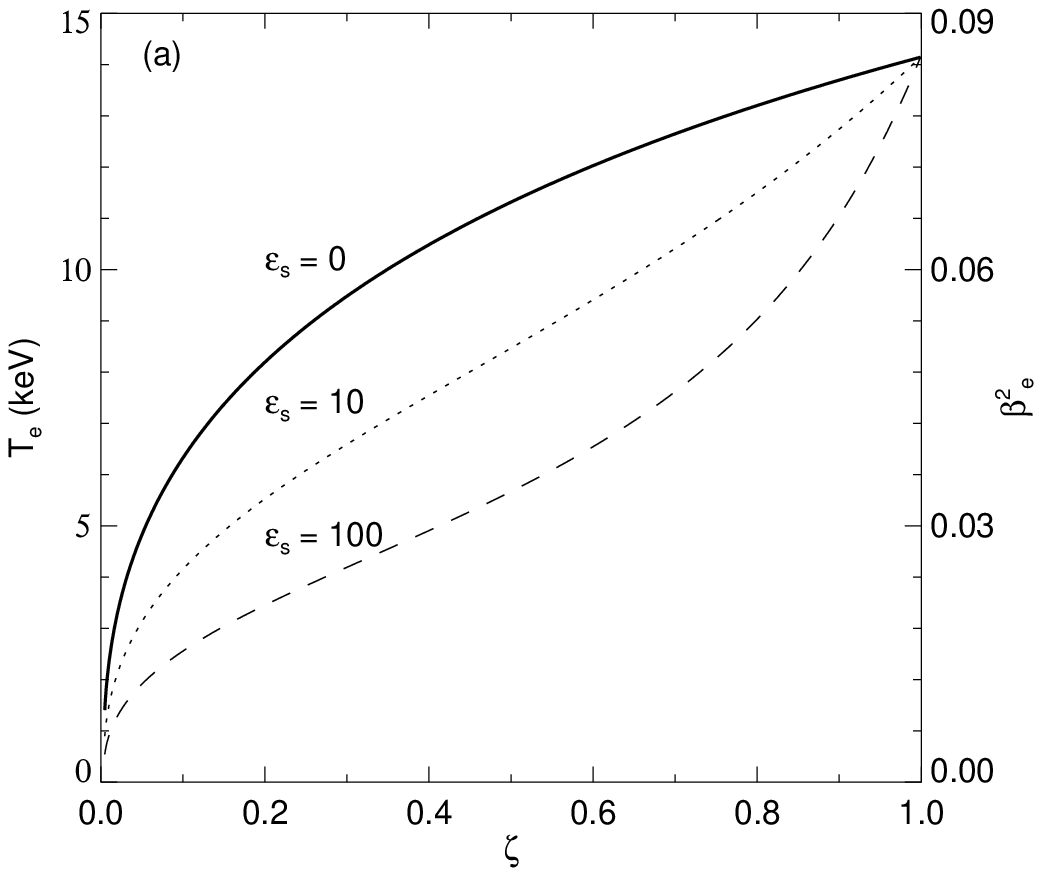}
 \includegraphics[width=5.5truecm]{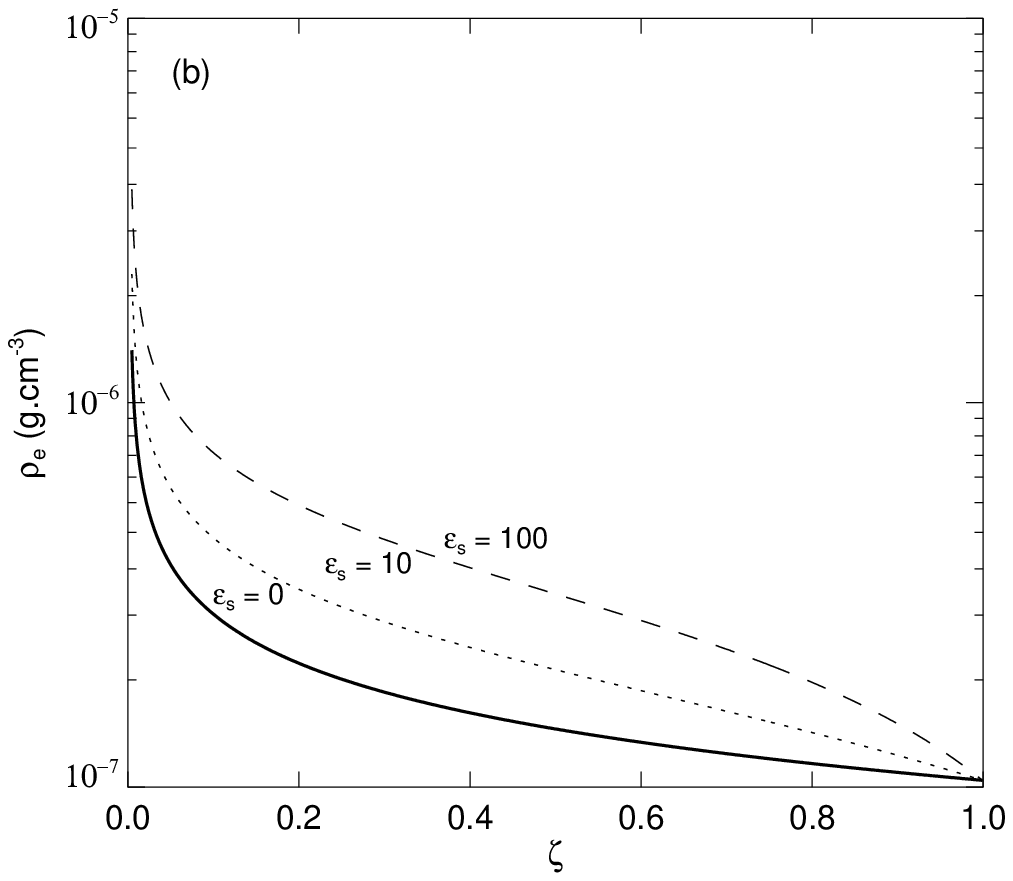}
 \includegraphics[width=5.5truecm]{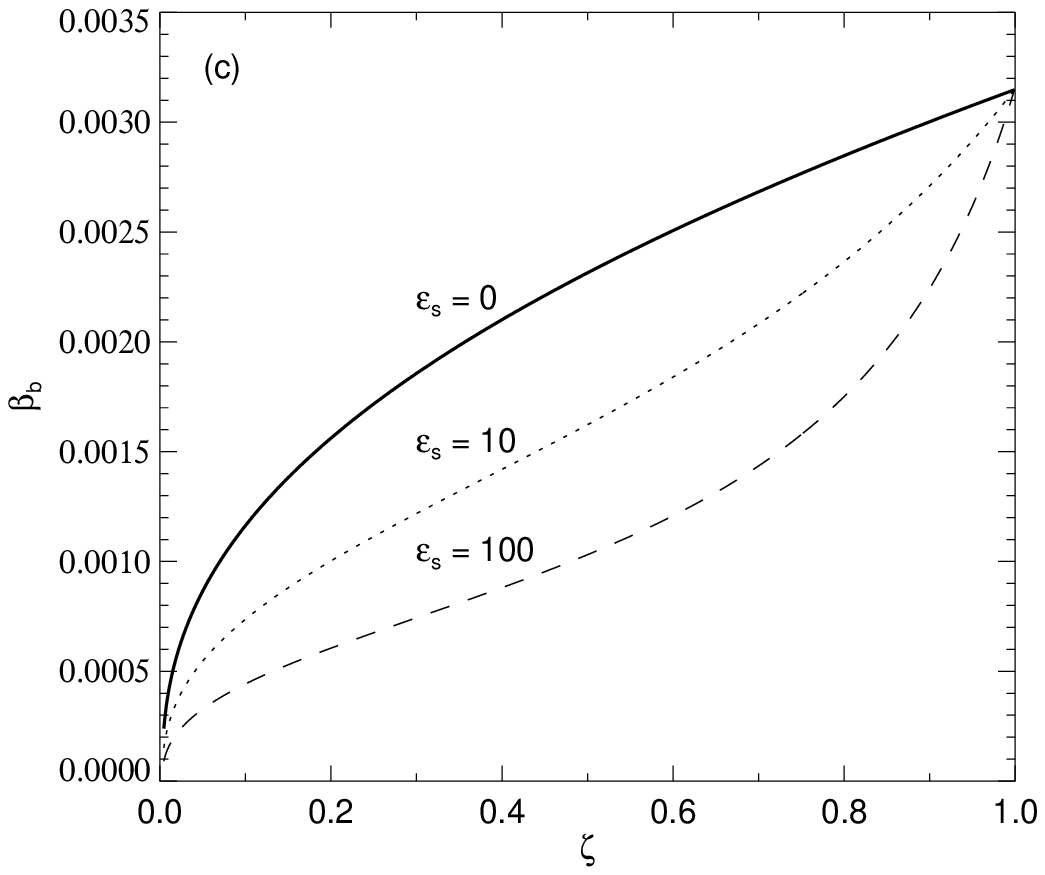}
  \caption{The profiles of (a) electron temperature $T_{\rm e}$, (b) mass 
    density $\rho_{\rm e}$ and (c) bulk velocity 
    $v_{\rm b} = \beta_{\rm b}c$ in the post-shock accretion column of an mCV 
    with $\Mwd = 0.5 \,\Msun$ and $\dot m = 10 \rm \, \gcms$. $\zeta = 0$ 
    corresponds to the base of the accretion column and $\zeta = 1$ to the 
    shock. The solid line shows the profile for the case where the ratio of 
    cyclotron to bremsstrahlung cooling at the shock is $\epsilon_{\rm s} = 0$,
    the dotted line for $\epsilon_{\rm s} = 10$ and the dashed line for 
    $\epsilon_{\rm s} =100$. In (a) the corresponding mean thermal electron 
    velocity $\beta_{\rm e} = v_{\rm e}/c$ is also shown as a function of 
    $\zeta$.}\label{profiles}
\end{figure*}

We present simulations of Compton scattering of Fe K$\alpha$ lines in an mCV 
accretion column for two different WD mass-radius values: 
$\Mwd = 0.5 \, \Msun, \, \Rwd = 9.2 \times10^8 \rm \, cm$ and 
$\Mwd = 1.0 \, \Msun, \, \Rwd = 5.5 \times10^8 \rm \, cm$ \citep*{Nauenberg}. 
For each mass, we consider two different specific mass accretion rates, 
$\dot m = 1~\gcms$ and $\dot m = 10~\gcms$, and inclination angles 
$i = 0\degree$, $45\degree$ and $90\degree$ with $\pm5\degree$ range. The 
cross-sectional radius of the accretion column is fixed at $0.1\Rwd$. We 
investigate the effect of varying the column width in section 3.3.

Figure~$\ref{profiles}$ shows the temperature, velocity and density profiles
of an accretion column for an mCV with WD mass $M_{\rm wd} = 0.5 \, \Msun$ and 
$\dot m = 10~\gcms$, for cases where the ratio of cyclotron to bremsstrahlung 
cooling at the shock is $\epsilon_{\rm s} =$ 0, 10, and 100. The temperature 
decreases monotonically from a shock temperature $kT_{\rm s} \approx 14 \, 
\rm keV$ ($\zeta = 1$), to a small finite value at the base of the column 
($\zeta = 0$).The mass density in the post-shock region is determined by 
$\rho(\zeta) = 4\dot m/v_{\rm b}(\zeta)$ and is a minimum at the shock and 
reaches a maximum at the base of the column. The bulk velocity of the 
accreting material at the shock is $0.25\vff$ and decreases to zero at the base
of the column, where the plasma settles on the WD surface. 

Figures~\ref{M1} and \ref{M0.5} show the simulated line spectra for the case 
where cyclotron cooling is negligible and bremsstrahlung cooling dominates 
($\epsilon_{\rm s} = 0$). Figure~\ref{cyclotron} shows the simulated profiles 
for $\Mwd = 0.5 \, \Msun$ and $\Mwd = 1.0 \, \Msun$ when cyclotron cooling 
dominates bremsstrahlung ($\epsilon_{\rm s} = 10$). In these cases, the 
neutral Fe K$\alpha$ line is emitted at the WD surface ($\zeta = 0$), where 
the bulk velocity is approximately zero and the thermal velocity of the plasma 
is small. The 6.675\,keV Fe K$\alpha$ line is emitted from the lowest few 
percent of the column ($\zeta \sim 0.003$) where the velocity of the infalling 
material and the thermal electron velocity are still relatively small. These 
lines thus show very little Doppler broadening. The 6.97\,keV line, however, 
is emitted much closer to the shock ($\zeta \sim 0.16$), in regions where the 
temperature of the accreting material is considerably higher and thus suffers 
more substantial Doppler broadening \citep[see][]{Wu01}.
The simulated line spectra for the case where the accretion column radius is 
fixed at $4.6 \times 10^{7} \, \rm cm$ is shown in Figure~\ref{constantradii}. 
Figure~\ref{emissivity} shows the simulated profiles for the case where the 
line photon injection along the flow is specified according to the emissivity 
profile model of \citet{Wu01}. In figure~\ref{GKPer} we compare our 
simulated line spectra with an observation of the mCV GK Per detected by 
\textit{Chandra}/HETGS.      

Table~\ref{FWHM} shows a comparison of the FWHM of the Fe K$\alpha$ lines for 
$\Mwd = 1.0 \, \Msun$ and $\Mwd = 0.5 \, \Msun$ with low and high $\dot m$ and
for $\epsilon_{\rm s} = 0$ and $\epsilon_{\rm s} = 10$.

\subsection{No Cyclotron Cooling}

\begin{figure*}{(a) $\Mwd = 1.0 \, \Msun$, $\dot m = 1~\gcms$, 
  $kT_{\rm s} = 34 \,   \rm keV$, no cyclotron cooling}\vspace{0.5cm}
  \includegraphics[width=18.0truecm]{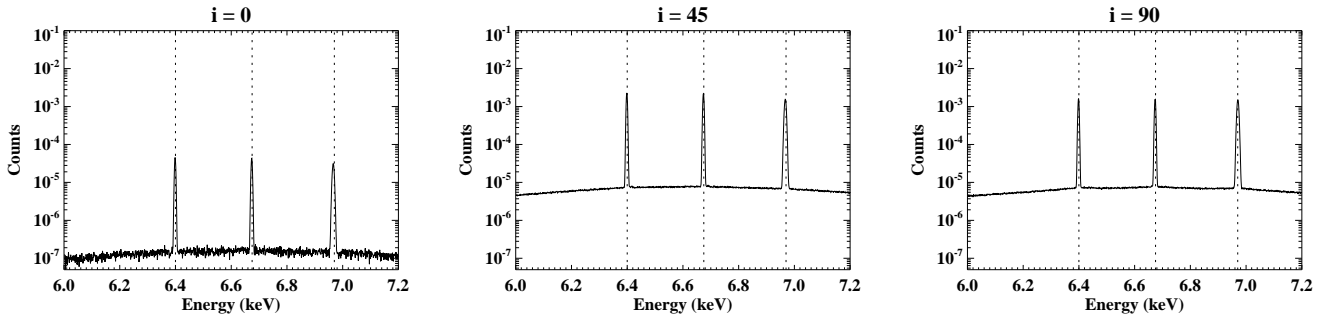}
  (b) $\Mwd = 1.0 \, \Msun$, $\dot m = 10~\gcms$, 
      $kT_{\rm s}= 46 \,\rm keV$, no cyclotron cooling\vspace{0.5cm}
  \includegraphics[width=18.0truecm]{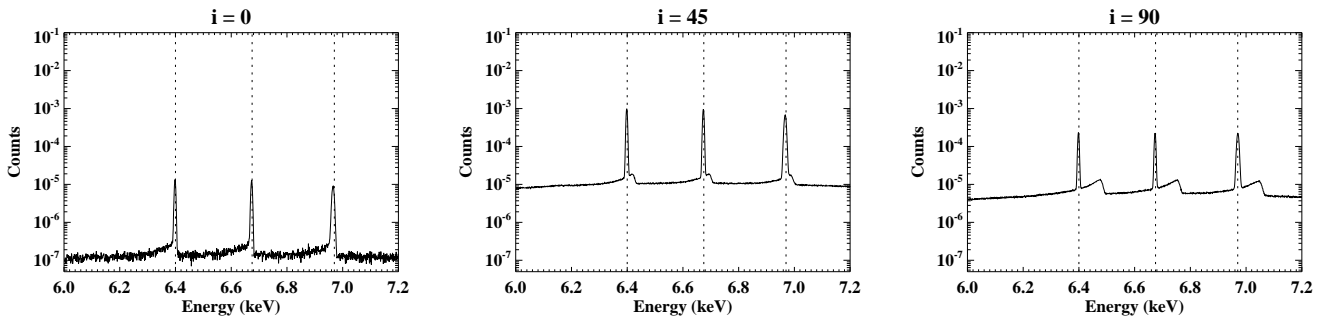}

    \caption{Profiles of Fe K$\alpha$ lines scattered by electrons in the 
    accretion column of an mCV with $\Mwd = 1.0 \, \Msun$ and with 
    (a) $\dot m = 1~\gcms$ and (b) $\dot m = 10~\gcms$. The shock temperature
    $kT_{\rm s}$ is indicated and the ratio of cyclotron 
    to bremsstrahlung cooling is $\epsilon_{\rm s} = 0$.}

\label{M1}
\end{figure*}

\begin{figure*}
  (a) $\Mwd = 0.5 \, \Msun$, $\dot m = 1~\gcms$, 
      $kT_{\rm s}= 14 \, \rm keV$, no cyclotron cooling\vspace{0.5cm}
  \includegraphics[width=18.0truecm]{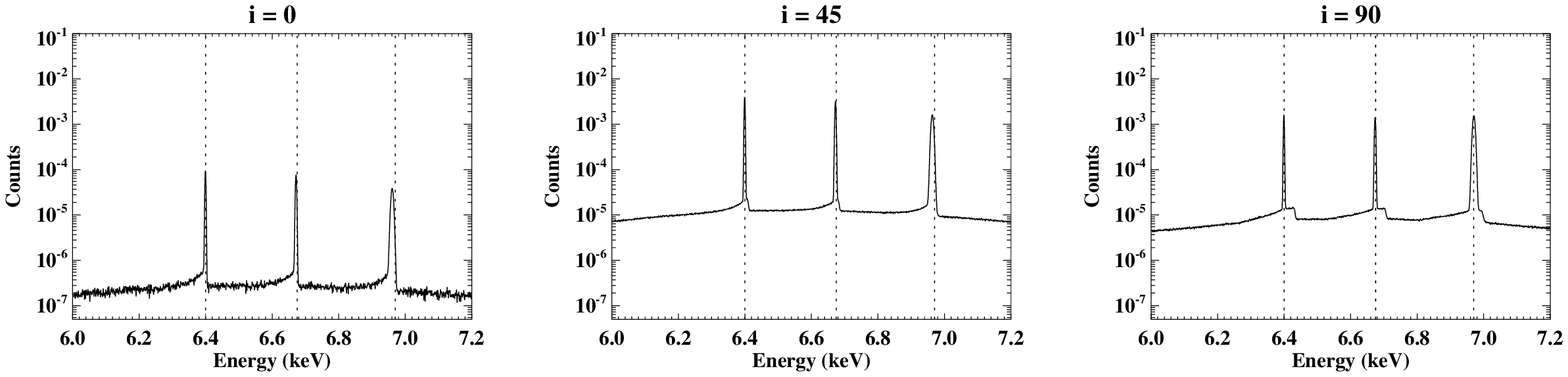}
  (b) $\Mwd = 0.5 \, \Msun$, $\dot m = 10~\gcms$, 
      $kT_{\rm s}= 14 \,\rm keV$, no cyclotron cooling\vspace{0.5cm}
  \includegraphics[width=18.0truecm]{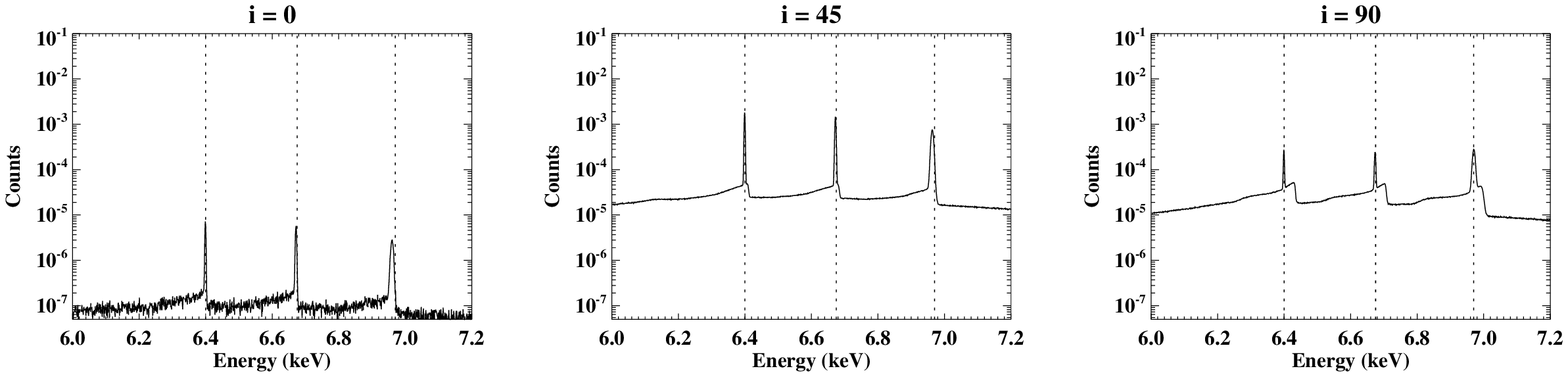}
    
  \caption{Profiles of Fe K$\alpha$ lines scattered by electrons in the 
    accretion column of an mCV with $\Mwd = 0.5 \, \Msun$ and with 
    (a) $\dot m = 1~\gcms$ (top panels) and (b) $\dot m = 10~\gcms$ (bottom 
    panels). 
    The shock temperature $kT_{\rm s}$ is indicated and the ratio of cyclotron 
    to bremsstrahlung cooling is $\epsilon_{\rm s} = 0$.}
\label{M0.5}
\end{figure*}

\begin{table}
  \centering
  \caption{FWHM (eV) values of the (a) 6.4\,keV, (b) 6.675\,keV and (c) 
           6.97\,keV lines calculated for two different values of $\dot m$ and 
	   for $M_{\rm wd} = 1.0 \, \Msun$ and $M_{\rm wd} = 0.5 \, \Msun$.
           The ratios of cyclotron to bremsstrahlung cooling at the shock 
	   are $\epsilon_{\rm s} = 0$ and $\epsilon_{\rm s} = 10$.}

\begin{tabular}{ccc}
\hline
(a)         &$M_{\rm wd} = 1.0 \, \rm \Msun$ &$M_{\rm wd} = 0.5\, \Msun$\\

6.4\,keV     & $\dot m = 1 \,(10)~\gcms$      & $\dot m = 1 \, (10)~\gcms$\\
\hline
$\epsilon_{\rm s}=0$  & $5 \, (7)$ & $4 \, (5)$\\
$\epsilon_{\rm_s}=10$ & $4 \, (5)$ & $3 \, (4)$\\
\hline
\hline
\end{tabular}

\begin{tabular}{ccc}
\hline
(b)           &$M_{\rm wd} = 1.0 \, \rm \Msun$ &$M_{\rm wd} = 0.5\, \Msun$\\

6.675\,keV     & $\dot m = 1 \, (10)~\gcms$ & $\dot m = 1 \, (10)~\gcms$\\
\hline
$\epsilon_{\rm s}=0$  & $6 \, (7)$ & $4 \, (5)$\\
$\epsilon_{\rm_s}=10$ & $5 \, (6)$ & $3 \, (5)$\\
\hline
\hline
\end{tabular}

\begin{tabular}{ccc}
\hline
(c)  & $M_{\rm wd} = 1.0 \, \rm \Msun$ &$M_{\rm wd} = 0.5\, \Msun$\\

6.97\,keV     & $\dot m = 1 \, (10)~\gcms$ & $\dot m = 1 \, (10)~\gcms$\\
\hline
$\epsilon_{\rm s}=0$  & $8 \, (9) $ & $10 \, (10)$\\
$\epsilon_{\rm_s}=10$ & $7 \, (15)$ & $8 \, (10)$\\
\hline
\hline
\end{tabular}
\label{FWHM}
\end{table}

When cyclotron cooling is negligible ($\epsilon_{\rm s} =0$), the 6.4, 6.675 
and 6.97\,keV lines are mostly emitted at heights above the WD surface of 
$x \approx 0$, $0$ and $1.3\times10^6 \, \rm cm$ for $\Mwd = 1.0 \,\Msun$ and 
$x \approx 0$, $1.0\times10^5$, and $5.1\times10^6 \, \rm cm$ for 
$\Mwd = 0.5 \,\Msun$, respectively \citep{Wu01}. Note that Doppler broadening 
is most significant for the 6.97\,keV line which is emitted further up in 
warmer regions of the post-shock column. Figures~$\ref{M1}$ and $\ref{M0.5}$ 
show that in all cases, there is substantial Compton broadening near the base 
of the line profiles due to scatterings in the post-shock region where photons 
may lose and gain energy. For example, for the 6.4~keV line emerging at 
$i=90\degree$ in Fig.~\ref{M1}a, the Compton broadened wings contain 
$\approx 28\%$ of the total line photons. The blue wing extends up to 
$\Delta E \approx$ 0.09~keV above the line centroid, while the red wing extends
down to $\Delta E \approx$ 0.16~keV (see Appendix). Compton features are 
generally more prominent at higher accretion rates since this gives higher 
optical depths in the column and hence, a higher scattering probability. For 
comparison, the average number of scatterings per photon in the $\dot m = 1 \, 
\gcms$ case is 0.5 but this increases to 2 in the $\dot m = 10 \, \gcms$ case. 

For $\Mwd = 0.5 \, \Msun$ (Fig.~\ref{M0.5}), a recoil tail redward of the line 
centroid can be seen in the cases $i = 45\degree$ and $90\degree$ specifically 
for high $\dot m$ (Fig.~\ref{M0.5}b) The shock temperature for these cases is 
$kT_{\rm s} \approx 14$ keV, while for the $\Mwd = 1.0 \, \Msun$ case, the 
shock temperature is much higher ($kT_{\rm s} \approx 34\rm \, keV$ for low 
$\dot m$ and $kT_{\rm s} \approx 46$ keV for high $\dot m$). The plasma 
temperature in the 1.0 $\Msun$ case is too high to produce any downscattering 
features and a large $\Mwd$ (small $\Rwd$) has an accretion column with a lower
optical depth, hence the absence of these features in Fig.~\ref{M1}. Again the 
downscattering features are more enhanced in the high $\dot m$ cases since the 
optical depth is higher (c.f. Figs.~\ref{M0.5}a and \ref{M0.5}b for 
$i = 90\degree$). Note also that the 6.97\,keV line centre is Doppler shifted 
slightly redward which is most noticeable for the 0.5 $\Msun$ case.  

In addition to the recoil signatures, upscattering features are also seen in 
the line spectra for the high $\dot m$ cases (Figs.~\ref{M1}b and \ref{M0.5}b) 
when $i = 90\degree$. For these inclination angles, photons emerge from the 
column with a final scattering angle cosine $\mu' \approx 0$. 
Equation~\ref{Compton} then gives 
\begin{equation}
  \frac{E'}{E} \approx 1 - \mu\beta \qquad.
\end{equation}
Since head on collisions ($\mu < 1$) have a higher probability, 
$E'\simeq (1 + |\mu|\beta)E$. The upscattering features are more prominent in
Fig.~\ref{M1}b than in Fig.~\ref{M0.5}b because the shock temperature, and 
hence $\beta$, is higher. The sharpness of the upscattering features seen in 
the 6.4\,keV profiles in particular can be attributed to the maximum possible 
energy gain when $\mu \approx -1$ and to very few hot electrons with energies 
beyond a few standard deviations of the mean thermal (Maxwellian) energy.

Scattering features are less prominent in the low $\dot m$ cases 
(Figs.~\ref{M1}a and \ref{M0.5}a), since the overall optical depth is smaller. 
For $\Mwd = 1.0 \, \Msun$ and $\dot m = 1~\gcms$ for instance, the optical 
depth across the column at the shock is $\tau \approx 0.04$, which is a factor 
10 smaller than that for the $\dot m = 10~\gcms$ case (since $\tau  \propto  
n_{\rm e} \propto \dot m$). For higher WD masses (i.e. smaller $\Rwd$), the 
optical depth across the column is smaller, so fewer photons are scattered, 
especially for low $\dot m$ cases. This explains the difference in scattering 
features in the profiles shown in Figs.~\ref{M1} and \ref{M0.5} (in particular,
the high $\dot m$, $i = 90\degree$ cases). 

Photons observed at an inclination angle of $i=0\degree$ propagate through the
entire length of the column before escaping. These photons propagate through a 
thick section of cold pre-shock flow and can thus downscatter, resulting in 
broadening redward of the line centre. However, the electron number density, 
and hence, optical depth, in the pre-shock flow is small so recoil effects 
are correspondingly small. Furthermore, fewer photons are detected at 
$i=0\degree$ because the solid angle centered around $i = 0\degree$ is 
considerably smaller than that centered around larger inclination angles.

\subsection[]{Cyclotron Cooling Dominated Flows}

\begin{figure*}
  \includegraphics[width=8.0truecm]{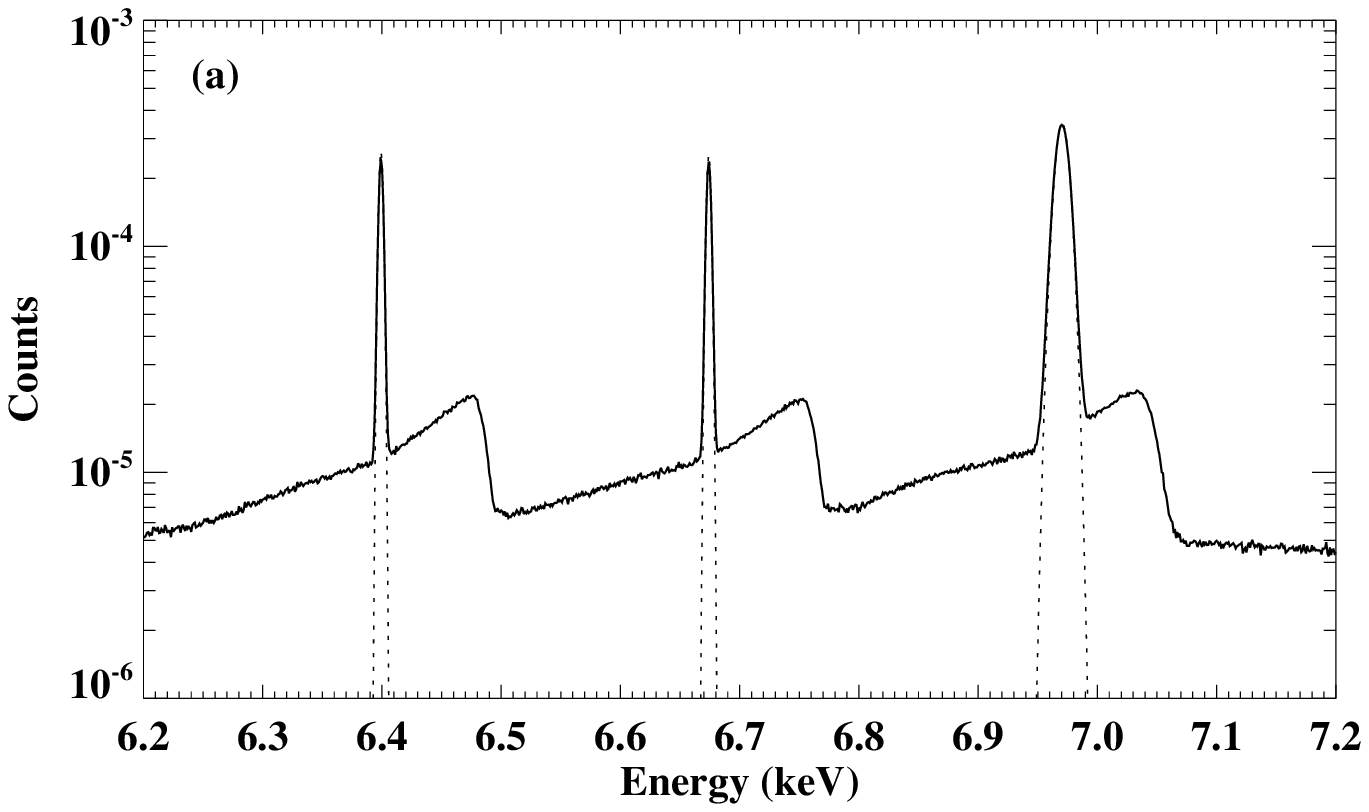}
  \includegraphics[width=8.0truecm]{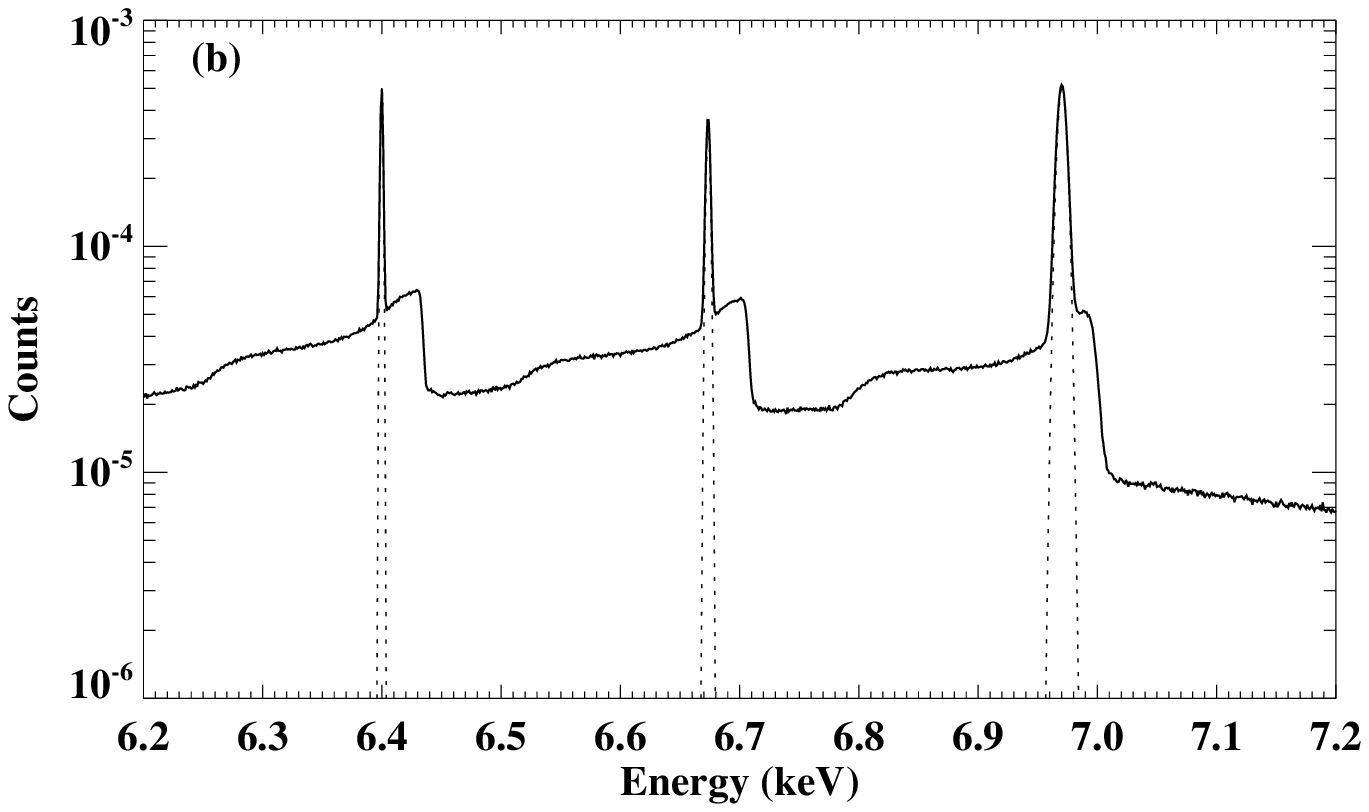}
  
 \caption{Simulated profiles of Compton scattered Fe K$\alpha$ lines in the 
    accretion column of an mCV with (a) $\Mwd = 1.0 \, \Msun$ and 
    (b) $\Mwd = 0.5 \, \Msun$. In both cases the inclination angle is 
    $i = 90\degree$, the specific mass accretion rate is $\dot m = 10~\gcms$ 
    and the ratio of cyclotron to bremsstrahlung cooling at the shock is 
    $\epsilon_{\rm s} = 10$. 
    The dotted curves show the contribution to each line profile by thermal 
    Doppler broadening only.}

\label{cyclotron}
\end{figure*}

Figure~$\ref{cyclotron}$ shows the profiles of Fe K$\alpha$ lines emitted from
mCVs with $\Mwd = 1.0\, \Msun$ and $\Mwd =0.5 \, \Msun$ where $i=90\degree$, 
$\dot m = 10~\gcms$ and the ratio of cyclotron to bremsstrahlung cooling at the
shock is $\epsilon_{\rm s} = 10$. Also plotted is the thermal Doppler 
broadened line profile before scattering (dotted curve). The additional cooling
of the plasma in the post-shock flow enhances the density and results in larger
scattering optical depths. Because the electron temperature of the shock-heated
region decreases as $\epsilon_{\rm s}$ increases, the peak emissivities of the 
6.675~keV and 6.97~keV Fe lines are found at larger 
values of $\zeta$, as can be seen in Fig.~\ref{profiles}, which shows the 
temperature, density and velocity profiles for different cases of 
$\epsilon_{\rm s}$. The additional cyclotron cooling results in a decrease in 
the electron temperature and bulk velocity in the post-shock column at a fixed 
$\zeta$. 

For $\epsilon_{\rm s} = 10$, the 6.4, 6.675 and 6.97\,keV lines are emitted 
from heights above the WD surface of $x \approx$ 0, $3.1\times10^5$ and 
$2.4\times10^7$ cm for $\Mwd = 0.5 \, \Msun$, and $x \approx$ 0, 0 and 
$5.3\times10^7$ cm for $\Mwd = 1.0 \, \Msun$, respectively \citep{Wu01}. The 
emission height of the 6.97\,keV line is significantly higher than in the 
$\epsilon_{\rm s} = 0$ case and consequently, thermal Doppler broadening is 
more prominent. The additional cooling in the accretion column also results in 
a lower plasma temperature at the base of the column, $\zeta = 0$ 
(see Fig.~\ref{profiles}), near where irradiation of neutral iron occurs and 
the fluorescent 6.4\,keV line is emitted. This line thus undergoes less thermal
Doppler broadening than in the $\epsilon_{\rm s}= 0$ case. As a result of 
Compton scattering, the line profiles show additional broadening and recoil 
tails redward of the line centre (Fig.~\ref{cyclotron}b). Upscattering features
are also evident in the 6.4 and 6.675\,keV lines especially for $\Mwd = 1.0 \, 
\Msun$ (Fig.~\ref{cyclotron}a).  

Overall, Compton scattering features are more prominent for cyclotron cooling 
dominated accretion columns, especially for high accretion rates. For photons 
viewed at $i = 90\degree$, upscattering and recoil features that are only 
marginally seen in the $\epsilon_{\rm s}=0$ case are considerably more 
pronounced in the $\epsilon = 10$ case. Thus, we expect Compton scattering 
features to be most conspicuous in Fe K$\alpha$ lines emitted in strongly 
magnetized mCVs accreting at a high rate.

\subsection{Effect of Accretion Column Radius}

\begin{figure*}
  \includegraphics[width=8.0truecm]{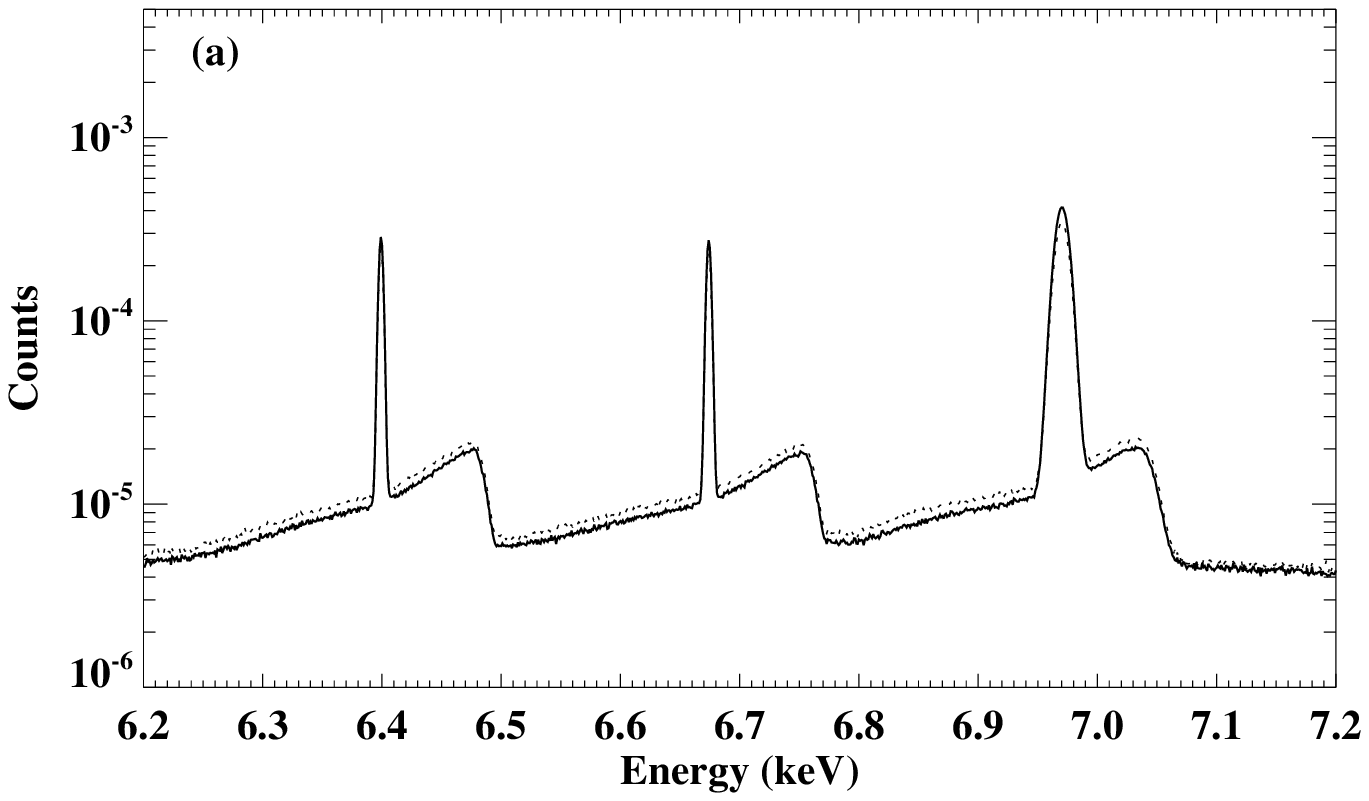}
  \includegraphics[width=8.0truecm]{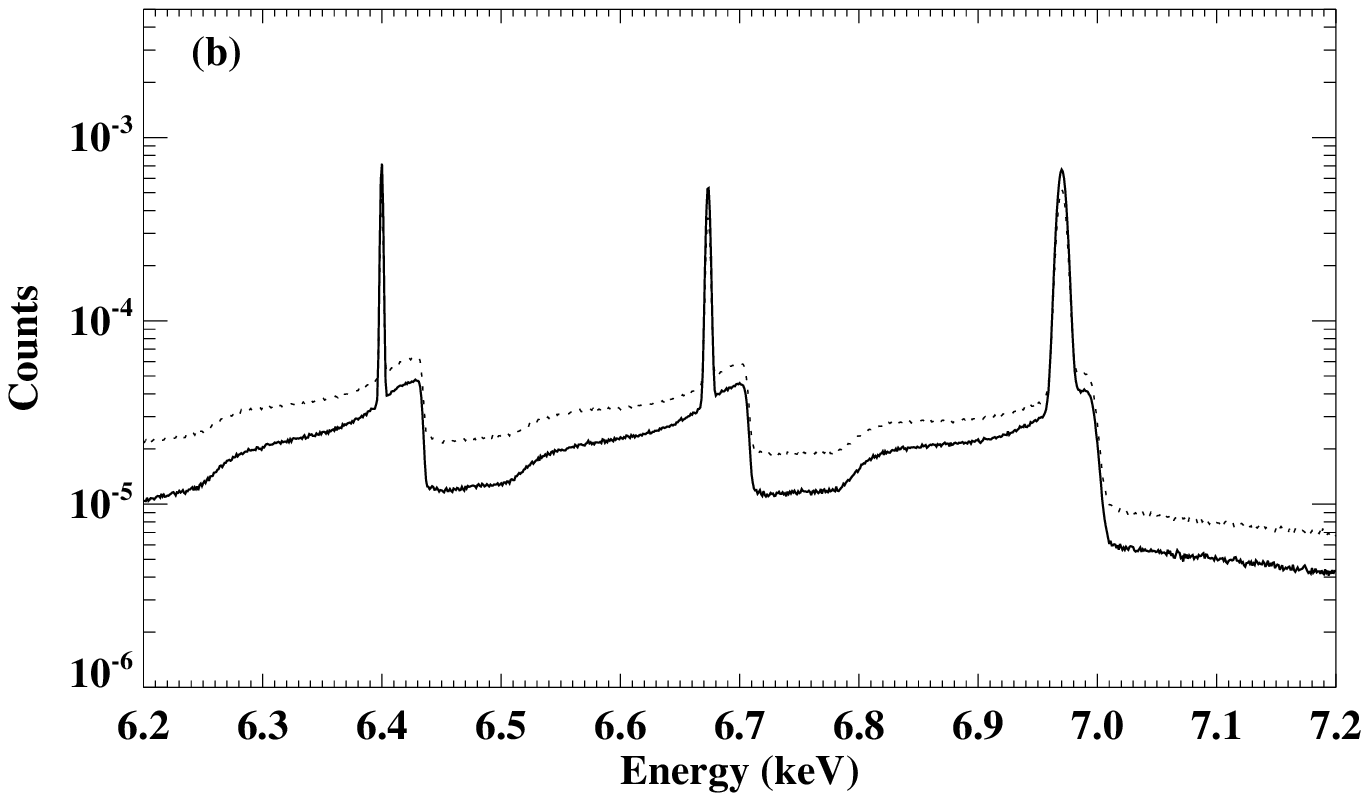}
  
 \caption{Simulated profiles of Compton scattered Fe K$\alpha$ lines in an 
    accretion column of an mCV with (a) $\Mwd = 1.0 \, \Msun$ and (b) 
    $\Mwd = 0.5 \, \Msun$  both with a specific mass accretion rate of 
    $\dot m = 10~\gcms$, a ratio of cyclotron to 
    bremsstrahlung cooling of $\epsilon_{\rm s} = 10$, an inclination 
    angle of  $i = 90\degree$ and accretion column radius of $4.6 \times 
    10^{7} \, \rm cm$ . The dotted curves show the profiles when the column
    radius is fixed at $0.1\Rwd$ (equivalent to the profiles shown in 
    Fig.\ref{cyclotron}).}

\label{constantradii}
\end{figure*}

The width of the accretion column in mCVs is poorly known and may vary 
significantly from system to system. In the results presented so far, we have 
fixed the accretion column radius to be $0.1\Rwd$. Here, we investigate the 
effect of relaxing this assumption. Figure~\ref{constantradii} shows the 
resulting profiles of Fe K$\alpha$ lines emitted from two mCVs of different 
masses ($\Mwd = 1.0 \, \Msun$ and $\Mwd = 0.5 \, \Msun$) but with the same 
absolute accretion column radius of $4.6 \times 10^{7} \, \rm cm$. This 
corresponds to $0.08\Rwd$ for the $1.0 \, \Msun$ case 
(Fig.~\ref{constantradii}a) and $0.05\Rwd$ for the $0.5 \, \Msun$ case 
(Fig.~\ref{constantradii}b). The other parameters used to produce the profiles 
are the same as those used for Fig.~\ref{cyclotron}, namely 
$\dot m = 10~\gcms$, $\epsilon_{\rm s} = 10$ and $i=90 \degree$. 
Fig.~\ref{constantradii} shows that the changes in the accretion column width 
generally result in changes near the base of the line profiles, which is 
broadened by multiple scatterings. The effect of increasing (decreasing) the 
width of the accretion column is to increase (decrease) the overall scattering
optical depth. 

The Thompson optical depth across the base of the accretion column is 
$\sim 20$ for the $1.0 \, \Msun$ case (Fig.~\ref{constantradii}a) and 
$\sim 40$ for the $0.5 \, \Msun$ case (Fig.~\ref{constantradii}b). In 
comparison, the Thompson optical depth across the column base for the same 
lines when the column has a width of $0.1\Rwd$ (dotted curves in 
Fig.~\ref{constantradii}) are $\sim 25$ for $\Mwd = 1.0\, \Msun$ and 
$\sim 80$ for $\Mwd = 0.5 \, \Msun$.  
Fig.~\ref{constantradii} indicates that due to the nonlinear nature of multiple
scatterings in the accretion column, particularly near the base, small changes 
in optical depth associated with the accretion column geometry can result in 
significant changes in line profiles. These changes mostly effect the base of 
the line profiles.

\subsection{Emissivity Profile Effects}
\begin{figure}
  \includegraphics[width=8.0truecm]{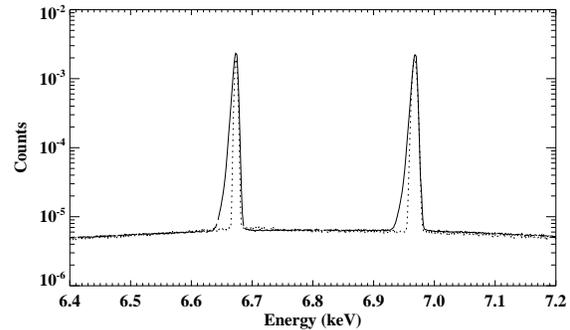}

 \caption{Simulated profiles of 6.675~keV and 6.97~keV Fe lines in an mCV 
   accretion column with $\Mwd = 1.0 \, \Msun$, a specific mass 
    accretion rate of $\dot m = 1~\gcms$, no cyclotron cooling and viewed at 
    an inclination angle of $i = 45\degree$ (solid lines). An emissivity 
    profile is used to disperse the photon injection site over a finite range 
    of heights along the post-shock column. The dotted curves show the 
    corresponding profiles for photons injected at a single height where the 
    emissivity peaks for the same mCV parameters (equivalent to the profiles 
    shown in Fig~\ref{M1}a, middle panel).}

\label{emissivity}
\end{figure}

Realistically, the photon source regions are determined by an emissivity 
profile along the post-shock region of the accretion column. In the results 
presented so far, all the photons were injected at a single height in the 
accretion column corresponding to the location of the peak in the emissivity
profile, as calculated by \citet{Wu01}. Here, we investigate the effect of 
spreading the photon injection site over a finite range of heights in the 
post-shock region according to the calculated emissivity profiles.
Only the 6.675 keV and the 6.97 keV line are studied in this manner since much 
of the fluorescence 6.4 keV yield derives from beneath the column, where X-ray 
irradiation is strongest so the injection site is always at the base of the 
column. Figure~\ref{emissivity} shows an example of the emissivity profile 
effect for the 6.675~keV and 6.97~keV Fe K$\alpha$ lines emitted in an mCV with
$\Mwd = 1.0 \, \Msun$, $\dot m = 1~\gcms$, $\epsilon_{\rm s} = 0$ and $i = 45 
\degree$. The dotted curves show the corresponding profiles for the case where 
photons are injected at a single height (where the emissivity peaks) for the 
same mCV parameters. These profiles are the same as those shown in 
Fig.~\ref{M1}a for the $i = 45 \degree$ case (middle panel).

The line profiles calculated using a realistic emissivity profile (solid curves
in Fig.~\ref{emissivity}) show some additional smearing, particularly redward 
of the line centre, compared to the single injection site case (dotted curves 
in Fig.~\ref{emissivity}). This can be attributed to a small fraction of 
photons now being emitted from regions closer to the shock, where the 
temperature is $\approx 34 \rm \,keV$ and the flow speed is 
$\approx 1230 \rm \, km.s^{-1}$. Thus, these photons are more affected by 
Doppler effects than other photons emitted from regions further away from the 
shock. The small smearing redward of the line centre is therefore due to a 
combination of Doppler shift and thermal broadening. The effect is not 
prominent because only a small fraction of photons are emitted close to the 
shock, as predicted by the emissivity profile \citep{Wu01}. The FWHM for the 
6.675~keV and 6.97~keV line is $\approx$ 8~eV and $\approx$ 9~eV respectively. 
This corresponds to an additional broadening of approximately 25\% for the 
6.675~keV line and 11\% for the 6.97~keV line. 

\subsection{Comparison with Observations}
\begin{figure*}
  \includegraphics[width=12.0truecm]{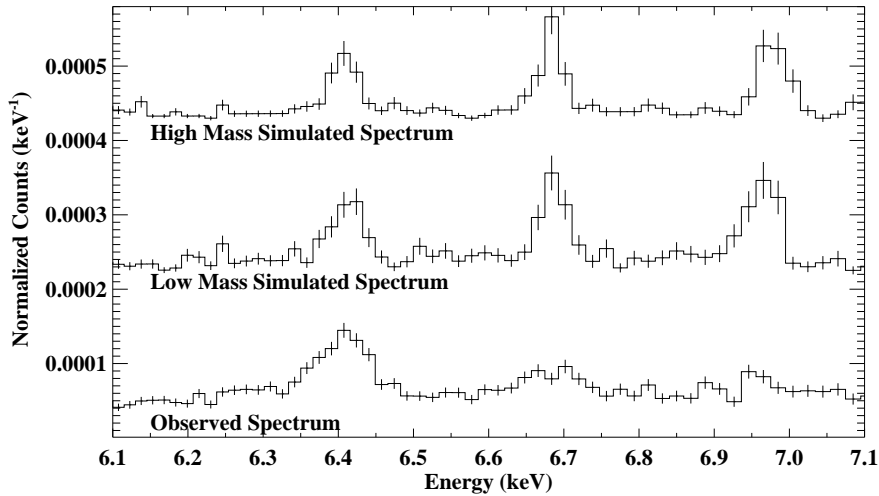}
  \caption{Fe K$\alpha$ emission lines for the mCV GK Per detected by 
           \textit{Chandra}/HETGS (bottom) and simulated by our model
	   for a lower and upper white dwarf mass $\Mwd = 0.63 \, \Msun$ 
	   (middle) and $\Mwd = 1.11 \, \Msun$ (top) both with a specific mass 
	   accretion rate $\dot m = 10~\gcms$.
	   The simulated spectra have been convolved with the 
	   \textit{Chandra}/HETGS response function. Cyclotron
           cooling is negligible and we have averaged the simulated lines
           over all inclination angles in an orbital period. An offset has been
	   added to the simulated plot to allow a comparison between the 
	   spectra. See text for further details.}
\label{GKPer}
\end{figure*}

\textit{Chandra}/HETGS observations of a number of mCVs were reported by
\citet{Hellier04}. The highest signal-to-noise spectrum was obtained in two 
observations of the mCV GK~Per during its 2002 outburst. This system is 
classified as an intermediate polar (IP); the WD accretor is thought to have a 
magnetic field of $\approx 1 \, \rm MG$ \citep{Hellier04}, and a mass of 
$\Mwd \geq 0.87 \pm 0.24 \, \Msun$ \citep{Morales}. Its typical X-ray 
luminosity in the $0.17 - 15$~keV bandpass is $L_{\rm x} \approx 7.14 \times 
10^{33} \, \rm erg \, s^{-1}$ \citep{Vrielmann}, which places a lower limit on 
the specific mass accretion rate, $\dot m \geq 7.3 \, \gcms$ (adopting 
$\Rwd \approx 6.7 \times 10^8 \, \rm cm$, inferred using the \citet{Nauenberg} 
mass-radius relation, and assuming that accretion proceeds onto a fraction 
$\approx 10^{-3}$ of the WD surface area; see e.g. \citealt{Frank}). 

Here we present a pilot study of GK Per and produce a simple comparison 
between the profiles of the simulated and observed spectra. A more detailed 
analysis of the data is left for future study.  
The summed first-order HETGS spectrum from the 2002 \textit{Chandra}\/ 
observations of GK Per is shown in Figure~\ref{GKPer}. The He-like and 
fluorescence Fe K$\alpha$ lines are detected, and there is a possible 
excess at the expected energy for the H-like line at 6.97\,keV. Also shown in 
Fig.~\ref{GKPer} are the upper and lower mass simulated HETGS spectra from a 
10~ks \textit{Chandra} observation of an mCV with parameters appropriate for 
GK Per: an upper mass of $\Mwd = 1.11 \, \Msun$ with $\Rwd = 5 \times 10^{8} 
\, \rm cm$, a lower mass of $\Mwd = 0.63 \, \Msun$ with $\Rwd = 8.2 \times 
10^8 \, \rm cm$ and for both masses $\dot m = 10 \, \gcms$ and $\epsilon_{\rm 
s} = 4 \times 10^{-5}$. We adopted inclination angles averaged 
over an orbital period of GK Per ($50\degree \leq i \leq 90\degree$) 
\citep{Hellier94}. We also added a normalized bremsstrahlung continuum with an 
electron temperature $kT_{\rm e} = 11 \, \rm keV$ \citep{Vrielmann}. 

The measured equivalent widths of the 6.4, 6.675 and 6.97\,keV lines in the 
\textit{Chandra}/HETGS spectra of GK Per are 260, 117 and 80 eV, respectively.
The relative strength of the lines in the simulated spectra is fixed by the 
number of photons used in each simulation and the assumed continuum flux level.
We used $10^8$ photons for all three lines, so that their relative strengths 
are comparable. We intend to relax this condition in future work. There is a 
remarkable similarity between the observed and the simulated spectra for the 
6.4~keV line in particular. Realistically, the actual strengths of the 6.675 
and 6.97\,keV lines depend on the ionisation structure of the flow 
\citep{Wu01}, since this determines the emissivity of the lines. The H-like and
He-like lines are thus potentially powerful diagnostic tools that can be used 
to deduce the physical properties of the post-shock accretion column in mCVs. 
The 6.4\,keV line strength, on the other hand, depends on the flux of the 
illuminating X-rays, which is largest near the WD surface. Fluorescent lines 
may also be produced in the surface atmospheric region around the accretion 
column \citep{Hellier04}, and this may contribute to the observed equivalent 
width of the 6.4\,keV line.

The observed fluorescent line in GK~Per exhibits a red wing extending to 
6.33~keV which \citet{Hellier04} attribute to Doppler shifts. In general we 
find that bulk Doppler shifts have a negligible effect on Fe K$\alpha$ lines. 
Our simulated spectra show a weaker red wing arising from downscatterings near 
the base of the accretion column. The observed spectrum also exhibits a 
shoulder extending 170~eV redward of the 6.4\,keV line centre, which 
\citet{Hellier04} suggest may be due to Compton downscattering. Our 
simulations confirm that recoil can indeed affect the fluorescent line when it 
is emitted at the base of the accretion column, although in our spectra 
(Fig.~\ref{GKPer}) there is no clear evidence that recoil extends down to 
170~eV redward of the 6.4~keV line.

\section{Summary and Conclusion}
We have investigated line distortion and broadening effects due to Compton 
scattering in the accretion column of mCVs using a nonlinear Monte Carlo 
technique that takes into account the nonuniform temperature, velocity and 
density profiles of the post-shock column. Scattered Fe K$\alpha$ lines were 
simulated for a range of different physical parameters: white dwarf mass, 
accretion rate, magnetic field strength and inclination angle. The photon 
source regions in the post-shock flow were determined from the ionization 
structure and the effects due to the bulk velocity Doppler shift and thermal
broadening are also considered. 

We find that line profiles are most affected by Compton scattering in the cases
of low white dwarf mass, high specific mass accretion rate, strong cyclotron 
cooling and oblique inclination angles. Both a lower white dwarf mass (or 
equivalently a larger white dwarf radius) and a higher accretion rate result in
a higher optical depth, hence more pronounced scattering effects. Strong 
cyclotron cooling associated with a high magnetic field strength results in a 
higher density in the post-shock flow and hence, a correspondingly larger 
optical depth. Recoil signatures are evident for oblique viewing angles 
when cyclotron cooling becomes important. These are due to scatterings near the
base of the flow, where both the dynamical and thermal velocities are small.
Sharp upscattering features are also seen in the line profiles for large 
viewing angles. These upscattering features are attributed to head-on 
collisions with hot electrons near the shock.  Photons emitted close to the 
shock in the accretion column display more thermal Doppler broadening. This is 
generally the case for the Fe~{\scriptsize{XXVI}} 6.97\,keV line. The Fe 
6.4\,keV and Fe~{\scriptsize{XXV}} 6.675\,keV lines are generally emitted at 
or near the base of the accretion column respectively, and show a relatively 
small amount of thermal Doppler broadening. Bulk velocity induced Doppler 
shifts are negligible compared with scattering.

We also investigated the effects of dispersing the 6.675 and 6.97~keV line 
photons using an emissivity calculated self-consistently from the 
ionization structure of the post-shock column. We find that, compared to 
injecting the line photons at a single height where the emissivity peaks,
using an emissivity profile does not significantly change the overall profiles
of the 6.675 and 6.97~keV lines. The resulting line profiles show a small 
degree of smearing redward of the line center due to additional dynamical and 
thermal Doppler effects associated with a small fraction of line photons
emitted close to the shock. The effect is most important for the 6.675~keV 
line. We estimate a fractional increase in the FWHM of no more than 
$\approx$~25\%.

In general, our simulations predict that recoil in scattering with cool 
electrons near the base of the column as well as upscattering by hot electrons 
near the shock can imprint signatures on the profile of lines emitted near the 
base of the flow. Such Compton signatures may thus be used to determine the 
primary source region of the fluorescent line. We predict that when 6.4\,keV 
photons are emitted in the dense, cool plasma at the base of the accretion 
column, they can suffer strong downscattering when propagating through the 
electrons streaming downward in the flow. The scattering probability is 
governed by the effective scattering optical depth, which increases with the 
mass accretion rate of the system. 

We convolved simulations with the \textit{Chandra} response function and 
compared them to \textit{Chandra}/HETGS observations of GK Per. Our results 
indicate that the red wing seen in the 6.4\,keV line in GK Per could be 
attributed to Compton recoil near the base of the flow. 

Finally we remark that tighter constraints on the dynamics and flow 
geometry in magnetized accreting compact objects can be obtained by considering
the polarization properties of the lines \citep[see][]{Sunyaev, Matt}. There is
currently considerable effort to develop X-ray polarimeters which can detect 
degrees of polarization of the order of one percent \citep{Costa}. The spectral
resolution of these detectors should be adequate to search for different 
polarization degrees in emission iron lines. It is possible to include a 
polarization treatment in our calculations, but the computational algorithm in 
our Monte Carlo code will need to be revised, which we leave for a future 
study.

\section*{Acknowledgments}
ALM thanks a University of Sydney Denison Scholarship. KW's visit to Sydney 
University was supported by a NSW State Expatriate Researcher Award. We thank 
an anonymous referee whose comments helped improve the paper considerably.

\appendix

\section{The width of the Compton shoulder}

Here, we derive an estimate for the width of the prominent wings near the base 
of the line profiles. Let the accretion flow be along the $z$-axis and consider
an observer on the $x-z$ plane. Let $i$ be the line-of-sight inclination angle 
of the accretion column, and let $\beta$ be the local velocity of the flow, 
normalized to the speed of light $c$, which can expressed as  
\begin{eqnarray} 
  \vec \beta & = & - \beta\ {\hat z} \  .  \nonumber 
\end{eqnarray}   
The normalized vector of a scattered photon of energy $E'$ propagating in the 
direction to the observer is 
\begin{eqnarray} 
  {\hat k}' & = & \sin i\ {\hat x}\ + \ \cos i \ {\hat z} \ . \nonumber
\end{eqnarray} 
Suppose that the normalized vector of the incident photon, with energy $E$, is 
\begin{eqnarray} 
  {\hat k} & = & \sin \theta \cos\phi\  {\hat x} \ + \ \sin\theta \sin \phi \ {\hat y} \ 
      + \ \cos\theta \ {\hat z} \ . \nonumber  
\end{eqnarray} 
Then the change in energy of the photon, after scattering with an 
electron in the flow, due to bulk motion is given by 
\begin{eqnarray}  
  \frac{E'}{E} & = & 
     \frac{ 1 - \mu \beta}
     {(1- {\mu}' \beta) + \frac{E}{\gamma m_{\rm e} c^2} (1 - {\hat k} \cdot {\hat k}' )} \ ,  \nonumber 
\end{eqnarray}    
where $\gamma = (1-\beta^2)^{-1/2}$, $\mu \beta = {\hat k} \cdot {\vec \beta}$, and ${\mu}' \beta = {\hat k}' \cdot {\vec \beta}$\ .  
In terms of the viewing inclination angle and the photon propagation vectors, 
\begin{eqnarray}  
  \frac{E'}{E} & = & 
     \frac{ 1 +  \beta \cos \theta}
     {(1 +  \beta \cos i) + \frac{E}{\gamma m_{\rm e} c^2}
        (1 -\sin i \sin\theta\cos\phi - \cos i \cos \theta )} \ .   \nonumber 
\end{eqnarray}  
Then for $\beta \ll 1$ and $E \ll m_{\rm e} c^2$,   
\begin{eqnarray}  
 \frac{E'}{E} & \approx &  
   1 +  \beta ( \cos \theta - \cos i )   
     - \lambda
        (1 -\sin i \sin\theta\cos\phi - \cos i \cos \theta ) \ .  \nonumber 
\end{eqnarray} 
It follows that 
\begin{eqnarray} 
  \frac{\Delta E}{E} & \approx &  \beta ( \cos \theta - \cos i )   
     - \lambda  (1 -\sin i \sin\theta\cos\phi - \cos i \cos \theta ) \ ,  \nonumber 
\end{eqnarray}  
where $\Delta E = E' - E$ and $\lambda = E/ m_{\rm e} c^2$.
  
The maximum energy downshift is caused by the recoil process when the photons 
are scattered by ``cold'' electrons (with $\beta \approx 0$). This occurs 
when $\phi = \pi$, $\theta = \pi - i$,  which yields  
\begin{eqnarray}
 \frac{\Delta E}{E} \bigg\vert_{\rm max~down }& =  &  - 2 \lambda \nonumber  \ . 
\end{eqnarray}  
This result is practically independent of the WD mass and the viewing 
inclination angle.  
For the 6.4~keV line, $\Delta E/E\vert_{\rm max~down}  \approx  - 2.50 \times 10^{-2}$.
 
The maximum energy upshift is caused by a Doppler shift when the photons are 
scattered by the fastest available downstream electrons (i.e. $\beta$ is no 
longer a negligible factor). The condition for its occurrence can be derived as
follows. Set  
\begin{eqnarray} 
  \frac{\partial}{\partial \phi} \left( \frac{\Delta E}{E}  \right) 
     & = & 0  \  ,      \nonumber 
\end{eqnarray} 
which gives two conditions for extrema with respect to the azimuthal 
coordinate: $\phi =0$ and $\phi = \pi$, corresponding respectively to 
\begin{eqnarray} 
   \frac{\Delta E}{E} & = & \beta (\cos \theta - \cos i) 
    - \lambda (1 \mp \sin i \sin \theta - \cos i \cos \theta) \ .  \nonumber 
\end{eqnarray} 
Differentiating the above expression with respect to $\theta$ and setting the 
resulting expression to zero yields the following condition for the extrema: 
\begin{eqnarray} 
  \frac{\sin \theta}{\cos \theta} & = & \pm \xi  \ , \nonumber 
\end{eqnarray} 
   where 
\begin{eqnarray}   
 \xi & = &    \frac{ \lambda \sin i}{\beta + \lambda \cos i}  \ . \nonumber 
\end{eqnarray}    
The first case ($\phi = 0$) leads to a maximum energy upshift, which requires
\begin{eqnarray} 
  \sin  \theta & = & \frac{1}{\sqrt{1+\xi^2}}    \nonumber \ ;  \\  \ 
  \cos \theta & = &  \frac{\xi}{\sqrt{1+\xi^2}}  \nonumber  \ .         
\end{eqnarray} 
At the viewing inclination angle $ i = \pi/2$, 
\begin{eqnarray} 
  \sin  \theta & = & \frac{\beta}{\sqrt{\beta^2+\lambda^2}}    \nonumber \ ; \\ 
  \cos \theta & = &  \frac{\lambda}{\sqrt{\beta^2+\lambda^2}}  \nonumber  \ .         
\end{eqnarray} 
Hence, the maximum energy upshift is given by 
\begin{eqnarray} 
 \frac{\Delta E}{E}\bigg\vert_{\rm max~up} 
   & \approx  & \sqrt{\beta^2 + \lambda^2} - \lambda \nonumber \ .   
\end{eqnarray}  
If we assume that the maximum $\beta$ takes the free-fall velocity at the 
WD surface, then $\beta$ is simply the reciprocal of the square root of the 
WD radius in the Schwarzschild unit, i.e. $\beta = \sqrt{2GM_{\rm wd}/R_{\rm wd}c^2}$. For $\Mwd = 1.0 \, \Msun$, $\beta = 2.31 \times 10^{-2}$, and for 
$\Mwd = 0.5 \, \Msun$, $\beta = 1.23 \times 10^{-2}$ (assuming the Nauenberg 
mass-radius relation \citep{Nauenberg}), this gives $\Delta E/E \vert_{\rm max~up} \approx 1.36 
\times 10^{-2}$ and $5.06 \times 10^{-3}$ respectively for the 6.4~keV line.  

The Compton shoulder of an Fe line is due to a single scattering event. 
For the 6.4~keV line, the broadening extends over $ 6.24 \rm \, keV \leq E 
\leq 6.49 \rm \, keV$ for a 1.0~$\Msun$ WD and $ 6.24 \rm \, keV \leq E \leq 
6.43 \rm \, keV$ for a 0.5~$\Msun$ WD (omitting thermal broadening and flow 
Doppler broadening), when viewed at $ i = \pi/2$.

\bsp

\label{lastpage}

\end{document}